\documentclass{article}
\usepackage{spconf,amsmath,graphicx,hyperref}
\DeclareMathOperator*{\argmin}{arg\,min}  % l'ho aggiunto (VERIFICA DI POTERLO FARE)
\usepackage{amsfonts,amssymb}      % gives \mathbb, \lVert, \Re, ...
\usepackage{subcaption} % fornisce environment subtable
\usepackage{caption}

\usepackage{array}
\usepackage{booktabs}
\usepackage[normalem]{ulem}
\usepackage{siunitx}   % per colonne numeriche compatte e allineate
\usepackage{adjustbox} % in preambolo
\usepackage[table]{xcolor}      % carica xcolor in modalità 'table'
\definecolor{HeaderGray}{gray}{0.92}   % tonalità identica (≈ 8 % nero)

\definecolor{HeaderGray}{gray}{0.92}  % colore di sfondo

\newcommand{\blockheader}[1]{%
  \rowcolor{HeaderGray}%
  \multicolumn{6}{c}{#1}\\ % 6 colonne, allineamento centrato
}

\title{EuleroDec: A Complex-Valued RVQ-VAE for Efficient and Robust Audio Coding}

\name{%
  Luca Cerovaz$^{1}$ \quad
  Michele Mancusi$^{2}$ \quad
  Emanuele Rodol\`a$^{1,3}$}

\address{$^{1}$ Sapienza University of Rome, $^{2}$ Moises Systems Inc.\\$^{3}$ Paradigma}
\begin{document}
\ninept
\maketitle
\begin{abstract}
Audio codecs power discrete music generative modelling, music streaming and immersive media by shrinking PCM audio to bandwidth-friendly bit-rates. Recent works have gravitated towards processing in the spectral domain; however, spectrogram-domains typically struggle with phase modeling which is naturally complex-valued. Most frequency-domain neural codecs, either disregard phase information or encode it as two separate real-valued channels, limiting spatial fidelity. This entails the need to introduce adversarial discriminators at the expense of convergence speed and training stability to compensate for the inadequate representation power of the audio signal. In this work we introduce an end-to-end complex-valued RVQ‑VAE audio codec that preserves magnitude–phase coupling across the entire analysis–quantization–synthesis pipeline and removes adversarial discriminators and diffusion post‑filters. Without GANs or diffusion we match or surpass much longer-trained baselines in-domain and reach SOTA out-of-domain performance. Compared to standard baselines that train for hundreds of thousands of steps, our model reducing training budget by an order of magnitude is markedly more compute-efficient while preserving high perceptual quality.

\end{abstract}
\section{Introduction}
\label{sec:intro}

Spectral-domain audio codecs decompose a signal into its time-frequency representation, typically via short-time Fourier transform (STFT). Although the magnitude (or power) spectrum carries most perceptual content, the phase spectrum is equally essential for faithful reconstruction: improper or discarded phase information leads to the smearing of transients, 'phasiness', and audible artifacts in the decoded signal. 
Many recent approaches, both in the waveform and spectral domains, employ multi-scale adversarial discriminators \cite{defossez2022high} \cite{kumar2023high} or score or flow-based post filters \cite{welker2025flowdecflowbasedfullbandgeneral} \cite{wu2024scoredec} \cite{10890277} to improve perceptual quality at the price of adversarial instability and slower convergence. An alternative line of work encodes complex spectra by splitting real-imaginary or magnitude–phase components and processing them independently \cite{wu2025complexdec,du2024apcodecspectrumcodingbasedhighfidelityhighcompressionrate,ai2024apcodecneuralaudiocodec,du2023funcodecfundamentalreproducibleintegrable}. However, this factorization neglects intrinsic phase–magnitude coupling and leads to parameter redundancy. Complex-valued neural networks (CVNNs) have shown promise in speech enhancement \cite{hu2020dccrndeepcomplexconvolution,halimeh2021complexvaluedspatialautoencodersmultichannel} and generative modeling \cite{nakashika20_interspeech}. Yet existing codec-oriented approaches either operate partially in the real domain \cite{Ru_2024}, breaking phase coherence, or are restricted to music and a single bitrate. To date, to the best of our knowledge, no codec realizes a fully end-to-end complex-valued analysis-quantization-synthesis pipeline without adversarial or post-filtering components.
In this work, we introduce the first fully end-to-end complex-valued neural audio codec for 24 kHz speech at 6 and 12 kbps. All stages of the pipeline - from waveform input to waveform reconstruction - are complex-valued and jointly optimized using Wirtinger calculus, without real-valued detours. The model integrates complex convolutions, attention, and normalization with 2×2 whitening, preserving the algebraic structure of the STFT and intrinsic amplitude–phase coupling. Without adversarial training or diffusion-based post-filters, the proposed approach achieves consistent in-domain and out-of-domain improvements at the target bitrates.
Our contributions are as follows:
(1) We propose the first end-to-end complex-valued VQ-VAE for audio, (2) our codec attains state-of-the-art quality without adversarial discriminators or score-based post-filters; (3) we reach and surpass state-of-the-art quality with fast, steady convergence and reduces training budget by $95\%$.

\section{Background and Related Work}
In this section, we give a brief introduction to Residual Vector Quantization \cite{lee2022autoregressiveimagegenerationusing} and complex-valued neural networks.
\subsection{Residual Vector Quantization}
In an audio codec based on an RVQ–VAE \cite{ai2024apcodecneuralaudiocodec} \cite{du2024apcodecspectrumcodingbasedhighfidelityhighcompressionrate} \cite{wu2025complexdec}, the encoder projects the complex STFT frame (or the waveform when working directly in the time-domain \cite{defossez2022high} \cite{borsos2023soundstormefficientparallelaudio}\cite{du2023funcodecfundamentalreproducibleintegrable} \cite{kumar2023high}) into a latent
vector $\mathbf z\!\in\!\mathbb{R}^{H}$.
A stack of $M$ residual codebooks
$\{\mathcal E^{(m)}\}_{m=1}^{M}$ iteratively approximates $\mathbf z$:
\[
  \mathbf r^{(m)} = \mathbf z - \!\!\sum_{j<m}\mathbf e_{k_j}^{(j)},\quad
  k_m = \argmin_k \bigl\lVert \mathbf r^{(m)}-\mathbf e_k^{(m)} \bigr\rVert_2 ,
\]
where $\mathbf e_{k}^{(m)}$ is the selected centroid at stage~$m$.
Only the index sequence $(k_1,\dots,k_M)$ is transmitted, yielding a
bit-rate $R_f\,M\,\log_2 K$. Because the latent space is real,
existing spectral codecs must split the complex STFT into two
real signals before quantisation. The first prevalent strategy \cite{ai2024apcodecneuralaudiocodec} \cite{du2024apcodecspectrumcodingbasedhighfidelityhighcompressionrate} \cite{feng2025stftcodechighfidelityaudiocompression} writes
the spectrum as modulus $|X|$ and unwrapped phase $\angle X$; two
independent RVQ pipelines are then trained, typically with a
log-power loss for magnitude and a wrapped angular loss for phase. 
\begin{figure}[!htb]
  \centering
  \includegraphics[width=1\columnwidth]{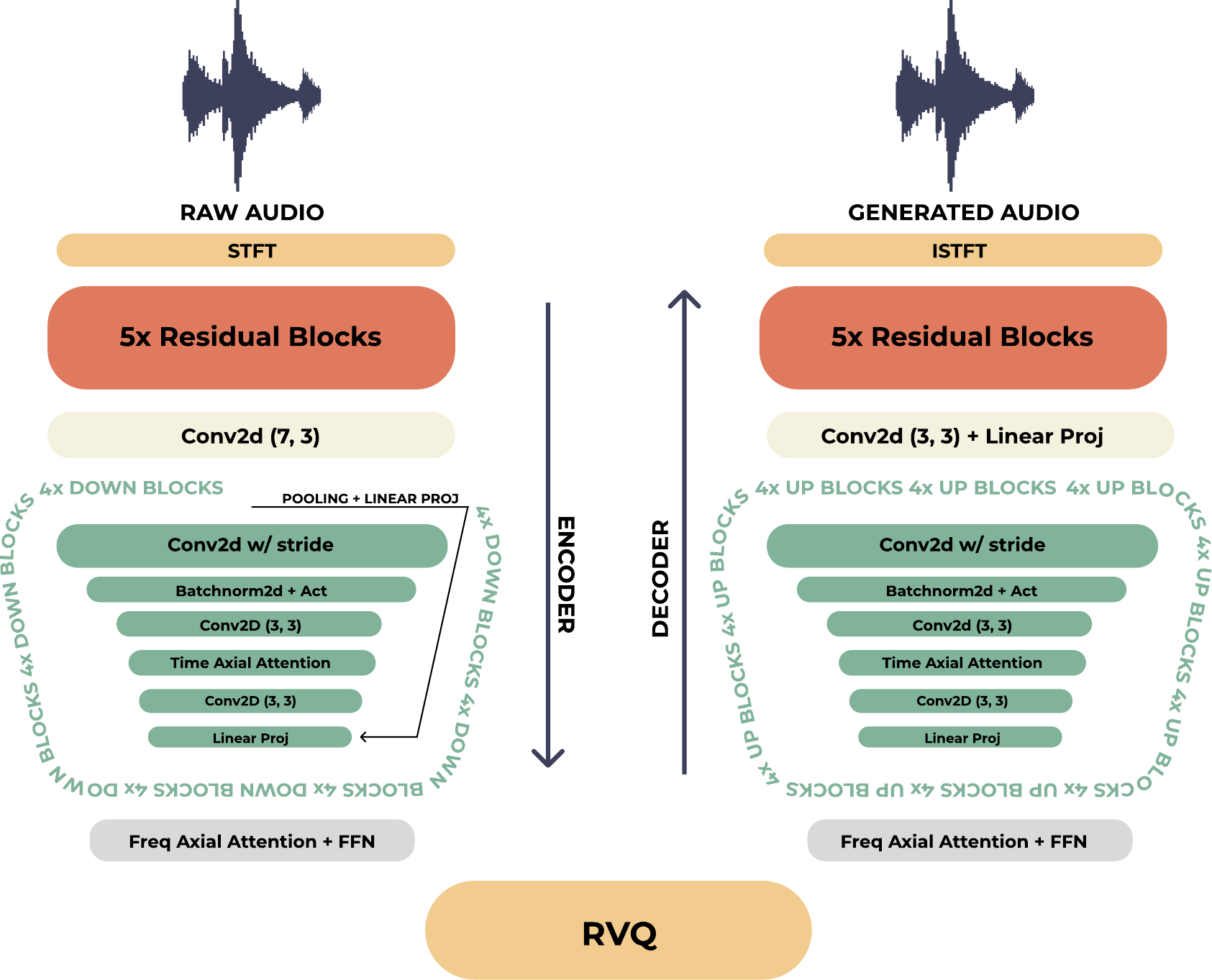}
  \caption{Model Architecture (All layers are complex-valued)}
  \label{fig:model_arch}
\end{figure}
The second strategy \cite{wu2025complexdec} \cite{feng2025stftcodechighfidelityaudiocompression} separates the spectrum into its real and imaginary components $\Re\{X\}$ and $\Im\{X\}$, feeds each map to an RVQ cascade optimised by Euclidean mean-squared error, and recombines the output as $\hat X=\hat R + j\hat I$ prior to the inverse STFT.
Both approaches, while duplicating the entire RVQ path into different streams, neglect the intrinsic correlation between magnitude and phase.

\subsection{Complex-valued neural networks}
Complex-valued neural networks \cite{DBLP:journals/corr/TrabelsiBSSSMRB17} represent inputs, weights, and activations as \(z = x + i y\) and compute with true complex algebra. A complex convolution is linear over \(\mathbb{C}\) and couples real and imaginary parts:
\[
\begin{aligned}
(w * z)[n] &= \sum_k w_k\, z_{n-k}, \\
w_k &= a_k + i b_k, \quad z_{n-k} = x_{n-k} + i y_{n-k}.
\end{aligned}
\]
This coupling lets the model learn amplitude-phase interactions rather than treating \(x\) and \(y\) as independent channels. A key property is phase equivariance \cite{DBLP:journals/corr/TrabelsiBSSSMRB17}: for any \(\phi\in\mathbb{R}\),
\[
f\!\left(e^{i\phi} z\right) \;=\; e^{i\phi} f(z),
\]
which preserves the geometry induced by \(U(1)\) rotations. Nonlinear layers are chosen to respect this symmetry. The \emph{modReLU} applies a threshold to the modulus, leaving the phase intact,
\[
\mathrm{modReLU}(z) \;=\; \mathrm{ReLU}(|z|+b)\,\frac{z}{|z|}.
\] 
\begin{figure}[!htb]
  \centering
  \includegraphics[width=0.61\columnwidth]{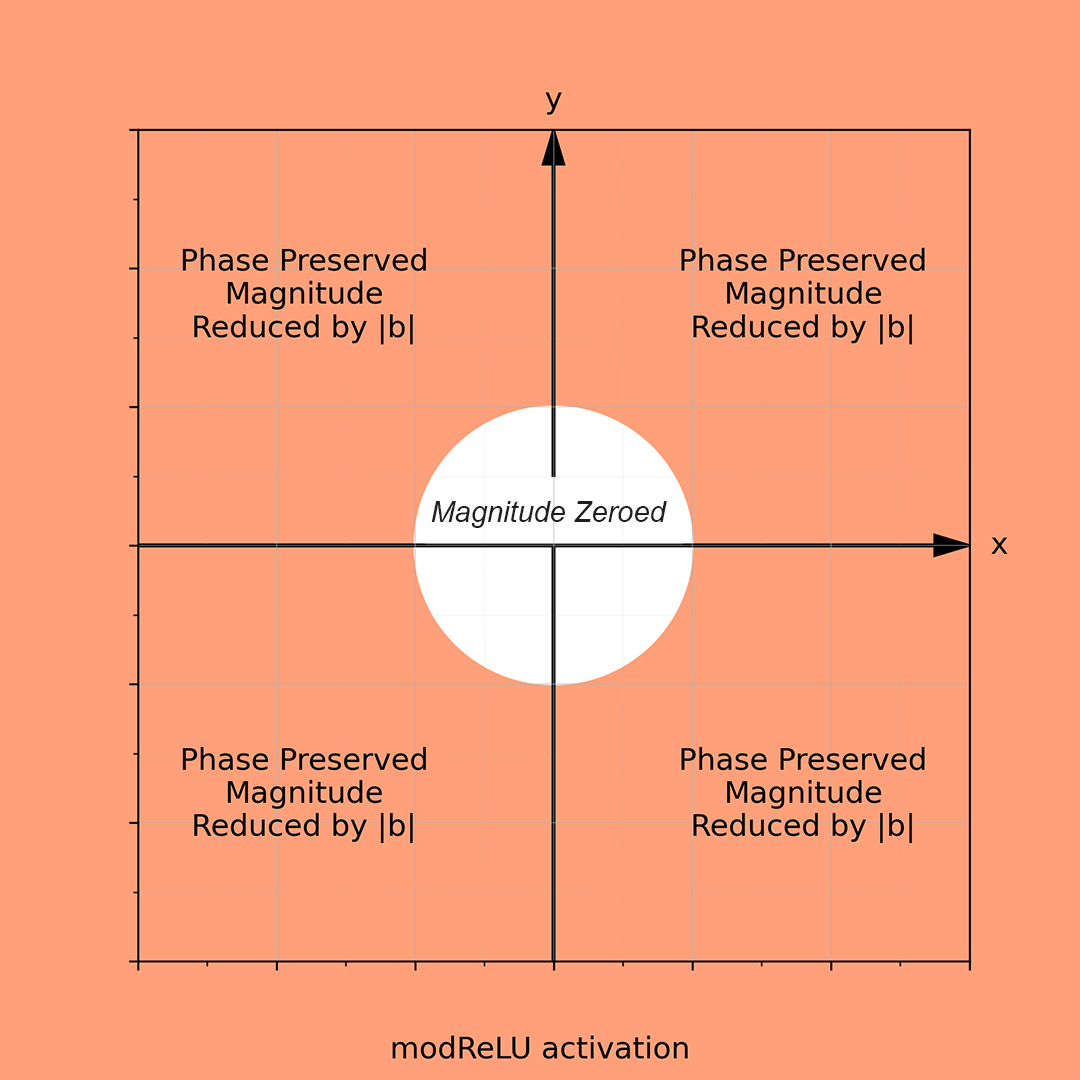}
  \caption{Visualization of the \textit{modReLU} activation: it applies a
threshold to the modulus, leaving the phase intact.}
  \label{fig:modrelu}
\end{figure}\\The effect of this function is in Figure.~\ref{fig:modrelu}
Normalization models cross-channel dependence by whitening \((x,y)\) jointly with a \(2\times2\) covariance, rather than normalizing parts separately. In audio coding, where phase carries timing and interference patterns, these properties make CVNNs a natural fit for spectrogram processing and improve generalization beyond the training domain \cite{Eilers_2023}.

\section{METHODS}We work directly in the complex domain with \\
\(x \in \mathbb{C}^{B \times C \times F \times T}\), of type \texttt{complex64}. 
We compute complex spectrograms with STFT at \(24\,\text{kHz}\) 
\((N_{\text{FFT}}=512,\ \text{win}=512,\ \text{hop}=64,\ \text{Hann})\) over 256 frames. 
For coding, we use RVQ with \(2048\)-entry codebooks.
At \(6\,\text{kbps}\), a temporal stride of \(8\) reduces 256 frames to 32 latent frames 
(\(\approx 46.9\ \text{tokens/s}\)). 
With fixed-length coding (\(11\ \text{bits/symbol}\) for \(K=2048\)) and \(12\) codebooks, 
this yields a bitrate of \(\approx 6.2\,\text{kbps}\).
At \(12\,\text{kbps}\), we use a temporal stride of \(4\), doubling the token rate to 
\(\approx 93.8\ \text{tokens/s}\) while keeping the same number of codebooks, 
resulting in a bitrate of \(\approx 12.4\,\text{kbps}\).
 As shown in ~\ref{fig:model_arch}, the architecture is a complex-valued VQ-VAE. All layers are complex-native: complex convolutions, complex normalizations (complex BatchNorm and complex RMSNorm), complex activations (ModReLU or GELU), and complex axial attention \cite{Eilers_2023}.
\subsection{Encoder and Decoder}
Four downsampling stages use the following anisotropic schedules (freq\(\times\)time). The decoder mirrors them with transposed convolutions and matching output padding.
\begin{center}
\scriptsize
\setlength{\tabcolsep}{5pt}
\renewcommand{\arraystretch}{0.95}
\begin{tabular}{@{}c c c c c@{}}
\toprule
Stage & $C_{\text{out}}$ & Kernel & Stride & Padding \\
\midrule
1 & 48  & (6,6) & (2,2) & (2,2) \\
2 & 64  & (6,1) & (2,1) & (2,0) \\
3 & 96  & (4,4) & (2,2) & (1,1) \\
4 & 128 & (4,4) & (2,2) & (1,1) \\
\bottomrule
\end{tabular}
\end{center} 
The encoder begins with five complex residual layers using dilations \cite{yu2017dilatedresidualnetworks} \(((1,1),\,(3,3),\,(3,5),\,(3,7),\,(1,1))\) to enlarge the receptive field while maintaining stable complex statistics. A complex \(3\times7\) convolution then prepares features for hierarchical compression. Four downsampling stages follow. In each stage, a gated skip branch computes adaptive complex average pooling \cite{DBLP:journals/corr/TrabelsiBSSSMRB17} of the input and applies a \(1\times1\) complex projection; this branch is summed with a main path that performs complex downsampling, complex normalization \cite{abdalla2023complexvaluedneuralnetworks}, complex activation, a \(3\times3\) complex convolution, complex axial self-attention \cite{Eilers_2023} along time, another \(3\times3\) complex convolution, and a final \(1\times1\) complex projection. The strided branch is summed with drop-path probability \(p=0.05\). 
\begin{figure}[!htb]
  \centering
  \includegraphics[width=0.70\columnwidth]{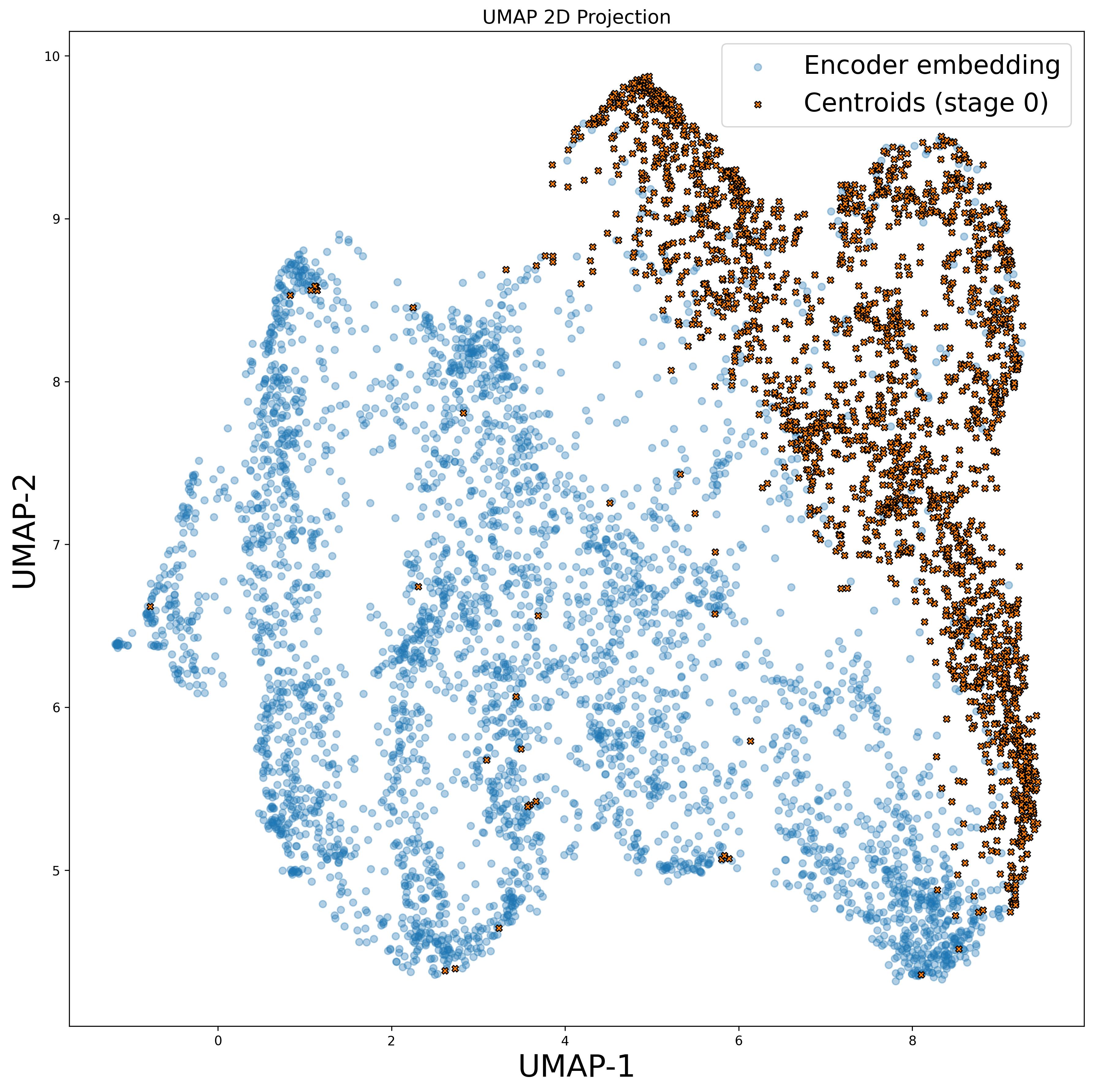}
  \caption{UMAP 2D of encoder embeddings and stage-0 centroids at 6 kbps. Our RVQ shows 100 \% code utilization and an effective perplexity of 73.2\% of the available codes, indicating broad, non-collapsed usage (a collapsed codebook would yield low perplexity). Distances are qualitative}
  \label{fig:umap2d}
\end{figure}
We keep the \(2\mathrm{D}\) spectrogram structure throughout the encoder so that frequency bins retain spatial relations rather than being collapsed into channels at early stages. Once spatial scales are compressed, the model applies axial attention across frequency and a complex feed-forward block before the quantizer. The decoder mirrors this mechanism without using the pooling branch, starting with frequency-axis attention and a complex feed-forward block, then four upsampling stages that invert the encoder ordering and restore the full-resolution complex spectrogram.
% --- Version 1

\subsection{Vector Quantizer}
We remain in the complex domain throughout quantization. The encoder output \(z_e \in \mathbb{C}^{B\times C\times F\times T}\) is reshaped by collapsing frequency into channels, yielding \(z_e^{\flat} \in \mathbb{C}^{B\times (C\!\cdot\!F)\times T}\). A complex linear projection \(W_{\text{in}}\in\mathbb{C}^{D\times (C\cdot F)}\) maps this merged representation to the code dimension. We then apply a residual vector quantizer with \(S\) stages \((S=12\text{ at }6\,\mathrm{kbps})\). Codebooks are initialized after a brief warm-up of 30 optimization steps by sampling centroid seeds from the current continuous encoder embeddings and adding small complex Gaussian noise, which increases diversity at start-up while keeping centroids near the data manifold.
At each stage, for every time index, vector quantization selects the nearest complex centroid under the Hermitian-induced Euclidean metric. Given \(\mathcal{E}=\{e_k\}_{k=1}^{K}\subset \mathbb{C}^{D}\) and \(x\in\mathbb{C}^{D}\),
\begin{align*}
d_k(x) 
       &= \lVert x\rVert_2^2 + \lVert e_k\rVert_2^2 - 2\,\mathrm{Re}\!\left(x^{\mathrm{H}} e_k\right), \\
k^\star(x) &= \arg\min_{k} d_k(x).
\end{align*}
The stage output accumulates a quantized reconstruction and updates the residual passed to the next stage. Encoder stability is promoted by a commitment loss that pulls \(z_e\) toward its assigned centroid without moving the codebook:
\[
\mathcal{L}_{\text{commit}}
\;=\; \beta \,\frac{1}{N}\sum_{n=1}^{N}
\left\|\, z_{e,n} - \operatorname{sg}\!\big(e_{k^\star(n)}\big) \right\|_2^2 .
\]
Where $sg$ is the stop gradient operation. Codebooks are updated via exponential moving averages of assignment counts and feature sums. To prevent early stickiness, we warm up the EMA \cite{Lancucki_2020} decay \(m\) across epochs, starting smaller ($0.98$) and annealing to the target value ($0.999$), which increases the effective codebook learning rate at the beginning and stabilizes it later.
\newcolumntype{T}{>{\ttfamily}l}
\begin{table}[t]
  \captionsetup{font=small}
  \centering
  \begin{adjustbox}{width=1\columnwidth,center}
  \scriptsize
  \setlength{\tabcolsep}{2.9pt}
  \renewcommand{\arraystretch}{1}
  \sisetup{
    table-number-alignment = center,
    table-figures-integer   = 2,
    table-figures-decimal   = 3
  }
  \begin{tabular}{@{}
      T
      S[table-format=2]
      S[table-format=1.2]
      S[table-format=1.2]
      S[table-format=5.2]
      S[table-format=1.3]
    @{}}
    \toprule
      & \multicolumn{1}{c}{ITERS}
      & \multicolumn{1}{c}{SI-SDR\,$\uparrow$}
      & \multicolumn{1}{c}{PESQ\,$\uparrow$}
      & \multicolumn{1}{c}{GDD\,$\downarrow$}
      & \multicolumn{1}{c}{ESTOI\,$\uparrow$} \\
    \midrule
    \blockheader{Out of Domain $24$ kHz (6 kbps)}\\
    \textbf{EuleroDec} & \textbf{35k} & \textbf{7.58} & \uline{2.16} & \uline{270} & 0.742 \\
    APCodec            & 700k & 0.35  & 1.91 & 596 & \uline{0.769} \\
    AudioDec           & 500k & -19.57 & 1.968  & \textbf{196} & 0.698 \\
    Encodec            & 500k & \uline{5.59} & \textbf{2.69} & 604  & \textbf{0.861} \\

    \toprule
      & \multicolumn{1}{c}{ITERS}
      & \multicolumn{1}{c}{SI-SDR\,$\uparrow$}
      & \multicolumn{1}{c}{PESQ\,$\uparrow$}
      & \multicolumn{1}{c}{GDD\,$\downarrow$}
      & \multicolumn{1}{c}{ESTOI\,$\uparrow$} \\
    \midrule
    \blockheader{In Domain $24$ kHz (6 kbps)}\\
    \textbf{EuleroDec} & \textbf{35k} & \textbf{10.5} & 2.47 & \uline{264} & 0.842 \\
    APCodec            & 700k & \uline{7.902} & \textbf{3.01} & 554  & \textbf{0.908} \\
    AudioDec           & 500k & -14.48 & 2.05  & \textbf{197} & 0.771 \\
    Encodec            & 500k & 7.47 & \uline{2.76} & 590 & \uline{0.905} \\

    \toprule
      & \multicolumn{1}{c}{ITERS}
      & \multicolumn{1}{c}{SI-SDR\,$\uparrow$}
      & \multicolumn{1}{c}{PESQ\,$\uparrow$}
      & \multicolumn{1}{c}{GDD\,$\downarrow$}
      & \multicolumn{1}{c}{ESTOI\,$\uparrow$} \\
    \midrule
    \blockheader{Out of Domain $24$ kHz (12 kbps)}\\
    \textbf{EuleroDec}   & \textbf{41k}  & \textbf{11.20}  & 2.57          & \textbf{257}  & 0.819 \\
    Encodec  & 500k & \uline{8.27}    & \textbf{3.63} & 591          & \textbf{0.925} \\
    APCodec  & 700k  & 5.63            & \uline{2.84}  & \uline{579}  & \uline{0.880} \\

    \toprule
      & \multicolumn{1}{c}{ITERS}
      & \multicolumn{1}{c}{SI-SDR\,$\uparrow$}
      & \multicolumn{1}{c}{PESQ\,$\uparrow$}
      & \multicolumn{1}{c}{GDD\,$\downarrow$}
      & \multicolumn{1}{c}{ESTOI\,$\uparrow$} \\
    \midrule
    \blockheader{In Domain $24$ kHz (12 kbps)}\\
    \textbf{EuleroDec}   & \textbf{41k}  & \textbf{13.67}  & 2.91         & \textbf{249}  & 0.901 \\
    Encodec  & 500k  & \uline{10.32}   & \textbf{3.77} & 578           & \textbf{0.953} \\
    APCodec  & 700k  & 5.93            & \uline{3.17}  & \uline{568}   & \uline{0.922} \\

    \bottomrule
  \end{tabular}
  \end{adjustbox}
  \caption{Evaluations of the codec's out-of-domain and in-domain robustness on the LibriTTS dataset. Best results are shown in \textbf{bold}, and second-best results are \uline{underlined}.}
  \label{tab:results-compact}
\end{table} This allows us to achieve $100\%$ of codes used throughout training (see fig ~\ref{fig:umap2d}). After the last stage, a complex linear map \(W_{\text{out}}\in\mathbb{C}^{(C\cdot F)\times D}\) projects back, and we unmerge frequency to recover \(z_q \in \mathbb{C}^{B\times C\times F\times T}\) for decoding.
We track per-code usage $u_k$ via an EMA of assignment counts; codes with $u_k \le \tau$ (dead-code threshold set as $0.9$) are flagged as dead.
For each dead code, with probability $p_{\mathrm{refresh}}=0.015$, we re-seed its vector from a randomly sampled feature $x_i$ in the current mini-batch (same group), optionally adding small complex Gaussian noise $\epsilon\!\sim\!\mathcal{CN}(0,\sigma^2 I)$ to encourage exploration. $\sigma$ is set as $0.001$
We then set $e_k\!\leftarrow\! x_i+\epsilon$, synchronize the EMA buffers $\bar e_k\!\leftarrow\! e_k$, and set $u_k\!\leftarrow\! \tau\!+\!1$ to avoid immediate re-pruning.

\section{EXPERIMENTS}
\subsection{Training Setup}
We train our codec on the \texttt{LibriTTS train.clean.100} subset (100h) \cite{koizumi2023librittsrrestoredmultispeakertexttospeech}. During training, we sample 0.680\,s segments at random for every epoch, ensuring that padding never exceeds 5\% so that cropped segments do not collapse at the tail of the waveform. Before feeding the model, we normalize the waveform and compute linear complex spectrograms, since log-magnitude features consistently degraded performance. Inference is performed by reconstructing with standard overlap-add across adjacent segments.
We train with the \texttt{AdamW} optimizer 
($\beta_1 = 0.9$, $\beta_2 = 0.99$, weight decay $= 7 \times 10^{-4}$) 
and a batch size of 16. The learning rate is set to $3 \times 10^{-4}$.
A linear warm-up is applied, followed by a cosine decay schedule with temporal warping, 
reducing the learning rate by a factor of 100 at convergence.  
We further use a multi-resolution mel L1 loss $\mathcal{L}_{\text{mel}}$, a multi-resolution spectrogram loss $\mathcal{L}_{\text{mrs}}$ combining spectral convergence and complex L1, and a quantization penalty $\mathcal{L}_q=\beta\,\mathcal{L}_{\text{commit}}$. The final objective is
\[
\boxed{
\mathcal{L}_{\text{total}}
=80w_{\text{mel}}\,\mathcal{L}_{\text{mel}}
+80w_{\text{cplx}}\,\mathcal{L}_{\text{gen}}
+50w_{\text{mrs}}\,\mathcal{L}_{\text{mrs}}
+0.1\,\mathcal{L}_q
}
\]
with $\beta=0.05$ for the commitment term.
\subsection{Experimental Setting}
We benchmark three strong neural codecs publicly available on the Libri TTS Corpus \cite{koizumi2023librittsrrestoredmultispeakertexttospeech} : \emph{AudioDec} \cite{Wu_2023} (streamable, time-domain, decoder is a HiFi-GAN–based vocoder \cite{kong2020hifigangenerativeadversarialnetworks}), \emph{EnCodec} \cite{defossez2022high}(streamable, time-domain), and \emph{APCodec} (not streamable, time-frequency-domain) \cite{ai2024apcodecneuralaudiocodec} at 6 and 12 kbps and 24 kHz. Similarly to APCodec, EuleroDec adopts a non-causal architecture to maximize phase coherence. While this precludes live streaming, the model remains highly efficient for high-throughput offline generation, achieving a Real-Time Factor (RTF) of 0.344 on an NVIDIA RTX 3090. We evaluate on two test splits: \emph{test-clean} (in-domain) and \emph{test-other} (out-of-domain). While the in-domain set mostly contains dry studio recordings, the out-of-domain set is characterized by mismatched microphones, reverberation, and background noise. This split is particularly valuable for assessing the robustness of the codec when distributional shifts occur. 
We train APCodec from scratch for $700.000$ iterations with batch size $16$ on the \texttt{LibriTTS train.clean.100} subset (100h), keeping the model architecture and quantizer as in the original work, but adapting the STFT front-end to 24 kHz by preserving the time spans of the analysis window and hop. For AudioDec we use the official public checkpoint at $500,000$ (only available at 6 kbps), for Encodec, we use a public community checkpoint released by zkniu on HuggingFace \cite{zkniu_encodec_pytorch_repo} trained for $500,000$ iterations with batch size 12. Both are trained on the Libri TTS Full Corpus. We test and train all codecs using an NVIDIA RTX 3090 GPU. All baselines rely on adversarial discriminators; EnCodec employs an MS-STFT discriminator, AudioDec adopts multi-period/related discriminators, and APCodec uses two discriminators (multi-period and multi-resolution). We reproduce official preprocessing and bitrates, and set an operating point near \(6\) and \(12\) kbps.
All models were trained until convergence, defined as no further loss improvements for at least three consecutive epochs. \subsection{Metrics}
Performance is assessed with four widely adopted metrics: 
SI-SDR is our primary waveform-based scale-invariant metric, directly sensitive not only to phase but also to residual noise (including quantization) and codec artifacts \cite{roux2018sdrhalfbakeddone}. PESQ (wideband) \cite{941023} measures perceptual quality, STOI (extended)  \cite{7539284} quantifies speech intelligibility,  and \emph{Group-Delay Distortion} (GDD) to quantify phase accuracy.
GDD measures error in the phase slope over frequency, which governs the temporal localization of transients. 
Low GDD indicates phase coherence and reduced pre-echo and temporal smearing. Together, SI-SDR, PESQ, STOI, and GDD cover amplitude fidelity, perceptual realism, intelligibility, and phase-driven timing.
\subsection{Evaluation}
Evaluation ~\ref{tab:results-compact} highlights a clear trade-off between in-domain perceptual tuning and out-of-domain robustness.
On LibriTTS clean at 6 kbps, APCodec optimizes PESQ and STOI, while EuleroDec secures the best SI-SDR and best (12 kbps) and second-best (6 kbps) in GDD, pointing to more faithful phase reconstruction. The contrast amplifies on LibriTTS other: APCodec's performance collapses, whereas EuleroDec improves relative ranking across all phase-aware metrics and remains competitive on PESQ. We attribute this to EuleroDec's end-to-end complex-valued design, which constrains representations to respect phase geometry and avoids adversarial objectives that can overfit content statistics. Notably, EuleroDec requires less than  ~$50,000$ training steps and uses no discriminator or diffusion post-filter, yet remains robust on phase-sensitive metrics. Complex-domain consistency appears to be the main driver of stable generalization and fast convergence at low bitrate. \subsection{Ablation: Time-Axial Attention
and Real vs Complex-Valued AE's}
We isolate the contribution of temporal axial attention by removing it from the Encoder and the Decoder. All settings (bitrate, data, optimizer, schedule, and seeds) are kept fixed ~\ref{tab:ablation-axial}. 
We report core quality metrics (SI–SDR, PESQ, STOI) and add encoder and decoder parameter count. This isolates the effect on long–range temporal coherence and phase timing while exposing quality-vs-compute trade-offs. We isolate the effect of complex-valued down/upsampling ~\ref{tab:ae} by comparing three mirroring autoencoders with 3 conv-activation-normalization blocks per side: a complex-valued AE ("cplx AE"), a real-valued AE that processes real and imaginary parts in separate channels ("split AE"), and a capacity-matched complex model with half hidden channels ("extra cplx") to counter the parameter inflation of complex layers. All models share identical data, segmentation, optimizer (AdamW), batch size, and loss: L2 on complex STFT coefficients plus complex spectral convergence with weight 0.2. The only architectural differences are complex convolutions, complex normalization with 2×2 whitening, and modReLU activations in the complex variants. Hidden widths are set so that parameter and compute budgets are comparable. Evaluation is performed on LibriTTS Clean with LSD (Log Spectral Distance) and PESQ. 
\begin{table}[t]
  \captionsetup{font=small} % >= 9 pt
  \centering
  \small                   % >= 9 pt (al posto di \scriptsize)
  \setlength{\tabcolsep}{3pt}
  \renewcommand{\arraystretch}{1}
  \begin{tabular}{lrrrr}
    \toprule
    Model & Params$\downarrow$ & SI\text{-}SDR$\uparrow$ & PESQ$\uparrow$ & STOI$\uparrow$  \\
    \midrule
    EuleroDec (full)   & 2{,}347{,}621 & \textbf{7.58} & \textbf{2.16} & \textbf{0.74}  \\
    \;-- w/o Time AxialAttn & \textbf{2{,}069{,}755} & 7.52 & 2.05 & 0.72 \\
    \bottomrule
  \end{tabular}
  \caption{Ablation of time–axial attention on 
\label{tab:ablation-axial}
\textit{test-other} at 24 kHz, 6 kbps}
\end{table}
\begin{table}[t]
  \captionsetup{font=small} % <-- riduce la caption solo qui
  \centering
  \begin{adjustbox}{width=0.7\columnwidth,center}
  \scriptsize
  \setlength{\tabcolsep}{3pt}
  \renewcommand{\arraystretch}{0.9}
  \sisetup{
    table-number-alignment = center,
    table-figures-integer   = 2,
    table-figures-decimal   = 3
  }
  \begin{tabular}{@{}
      T
      S[table-format=2]
      S[table-format=1.1]
      S[table-format=1.1]
      S[table-format=5.2]
      S[table-format=1.1]
    @{}}
    \toprule
      & \multicolumn{1}{c}{Hid. Dim.}
      & \multicolumn{1}{c}{LSD\,$\downarrow$}
      & \multicolumn{1}{c}{PESQ\,$\uparrow$} \\
    \midrule
    \blockheader{Evaluation of AE's}\\
    split AE & 36 & 0.72 & 2.06  \\
    extra cplx   & 16 & 0.58 & 2.60 &  &  \\
    cplx AE & 22 & 0.49 & 3.48 & \\

    \bottomrule
  \end{tabular}
  \end{adjustbox}
  \caption{Evaluations of a naive implementation of real and complex-valued AE on LibriTTS clean.}
  \label{tab:ae}
\end{table}
\section{CONCLUSION}
We presented \textit{EuleroDec}, the first end-to-end complex-valued RVQ-VAE audio codec. Operating entirely in $\mathbb{C}$, it preserves magnitude–phase geometry and attains strong perceptual quality with accurate phase reconstruction. At 6 and 12 kbps, it remains competitive in-domain and achieves state-of-the-art performance out-of-domain on LibriTTS-other on phase coherence and waveform fidelity, all without adversarial discriminators or score-based post-filters. Despite a lightweight schedule of less than ($50,000$ steps), the model is robust and sample-efficient, pointing to complex-domain consistency as a key driver of generalization. 
% References should be produced using the bibtex program from suitable
% BiBTeX files (here: strings, refs, manuals). The IEEEbib.bst bibliography
% style file from IEEE produces an unsorted bibliography list.
% -------------------------------------------------------------------------
\section*{\normalsize\bfseries ACKNOWLEDGMENTS}
This work is supported by the MUR FIS2 grant n. FIS-2023-00942 "NEXUS" (cup B53C25001030001), and partly by Sapienza University of Rome via the Seed of ERC grant "MINT.AI" (cup B83C25001040001).
\bibliographystyle{IEEEbib}
\bibliography{refs}

\end{document}